\documentclass[journal]{IEEEtran}
\usepackage{amsmath,amsfonts,amssymb}
\usepackage{array}
\usepackage{booktabs}
\usepackage{graphicx}
\usepackage{multirow}
\usepackage{makecell}
\usepackage{url}
\usepackage{balance}
\usepackage[caption=false,font=normalsize,labelfont=sf,textfont=sf]{subfig}
\usepackage{stfloats}
\usepackage{textcomp}

\begin{document}

\title{DiaHeads: Jointly Generating Coordinated 3D Dialogue Heads\\ with Non-verbal Cues from Dual Audio}

\author{Jiahe Wang, Yuxin Wang, Xin Yun, Binyan Chen, Tong Pan, Zhenjie Liu, and Shangfei Wang%
\thanks{The authors are with the School of Computer Science and Technology, University of Science and Technology of China, Hefei, Anhui, China (e-mail: wangjiahe317@mail.ustc.edu.cn; wyx2020@mail.ustc.edu.cn; yx9329@mail.ustc.edu.cn; sztn@mail.ustc.edu.cn; pantong@mail.ustc.edu.cn; liuzhenjie@mail.ustc.edu.cn; sfwang@ustc.edu.cn).}%
\thanks{Corresponding author: Shangfei Wang.}%
\thanks{Manuscript prepared for submission to the IEEE Transactions on Multimedia.}}

\maketitle

\begin{abstract}
Audio-driven 3D head animation has made considerable progress in generation realism, yet generating coordinated behavior for two-person dialogue remains difficult, because current dialogue head generation methods overlook the modeling of non-verbal behavior semantics between the two participants. To address this limitation, we present DiaHeads, a framework that generates the behavior of both participants jointly. The two audio streams are concatenated and decoded together, so each participant's motion is conditioned on the other, which makes the interaction part of the generation itself. We further introduce four discrete non-verbal cues, gaze-on, head shaking, nodding, and smiling, and fuse them with the continuous parameters in a shared latent space through a product-of-experts posterior. These cues carry the interaction semantics and constrain how the continuous motion is generated. A flow-based prior over the shared latent variable replaces the standard Gaussian prior and makes the latent motion temporally dependent. These designs therefore yield coordinated 3D dialogue heads. To support this task we build CANDOR-FG, a dialogue corpus derived from CANDOR and augmented with synchronized FLAME parameters and 3D gaze for both interlocutors. Experiments on CANDOR-FG show that DiaHeads achieves competitive per-participant fidelity and substantial gains in interpersonal coordination over single-speaker and dyadic baselines. In addition, ablations confirm that joint generation, the flow-based prior, and multi-cue supervision are the sources of this coordination. Finally, a user study shows that the advantage of DiaHeads comes from perceived interactional coordination rather than from raw motion fidelity alone.
\end{abstract}

\begin{IEEEkeywords}
3D Dialogue Heads Generation, Non-verbal Cues, Social Interaction Modeling.
\end{IEEEkeywords}

\section{Introduction}
\label{sec:intro}

\begin{figure}[!t]
  \centering
  \includegraphics[width=0.95\columnwidth]{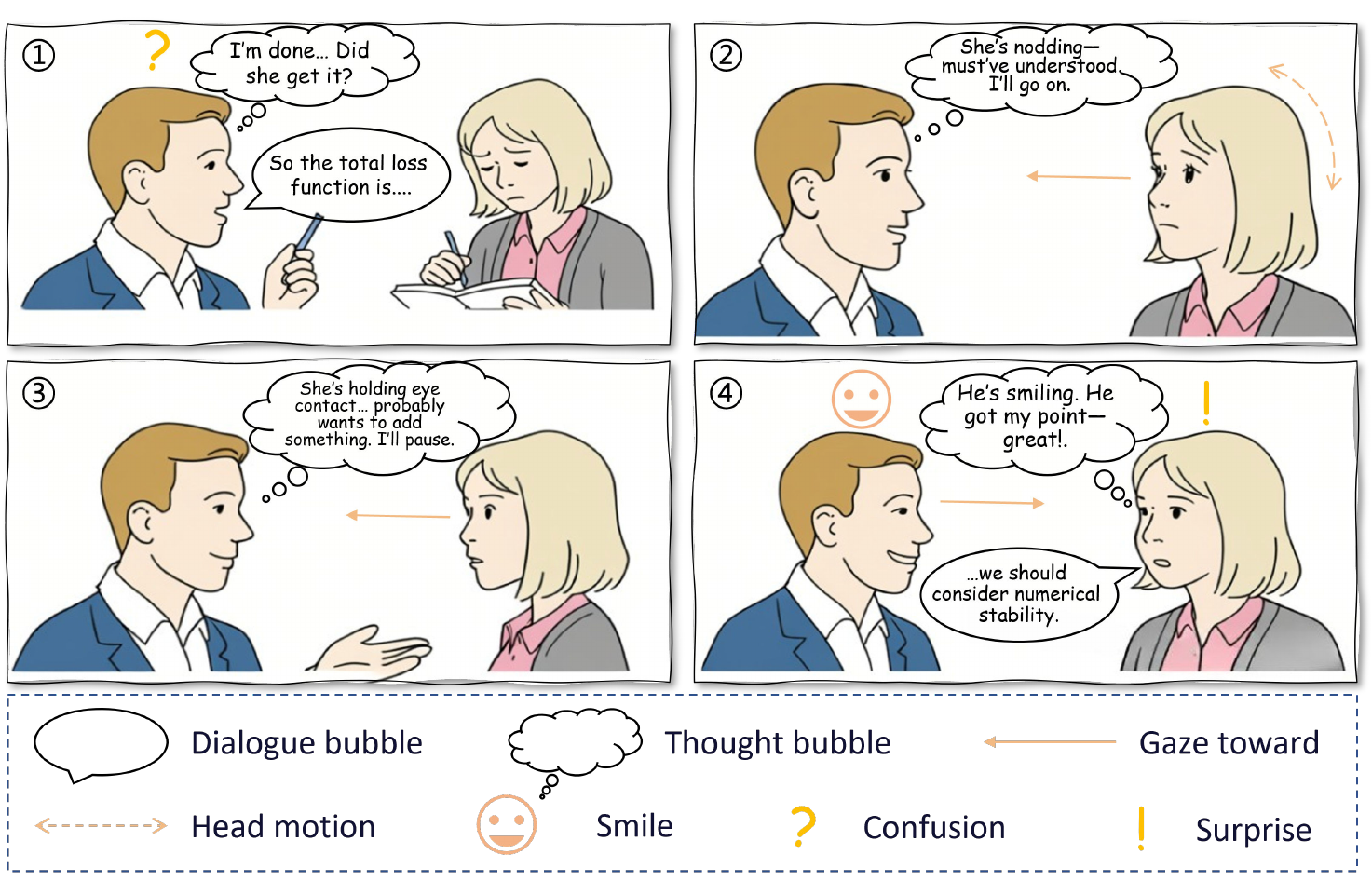}
\caption{Non-verbal cues as the ``social glue'' of a dyadic exchange. A is the person on the left and B is the person on the right. Panel 1 shows A speaking while awaiting feedback. Panel 2 shows B affirming with a nod and eye contact. Panel 3 shows B's gaze signaling a wish to add something while A pauses. Panel 4 shows B speaking while A smiles. This sequence illustrates how non-verbal cues regulate dialogue and sustain interpersonal coordination even without explicit verbal content.}
  \label{fig:cover}
\end{figure}

Generating realistic interactions between characters in dyadic dialogues supports immersive content creation, such as animating non-player-character conversations in games or synthesizing dialogue sequences for films. Beyond accurate lip synchronization, which has been extensively studied for single speakers \cite{cudeiro2019capture,fan2022faceformer,xing2023codetalker}, coordinated interaction requires the synthesis of a rich array of non-verbal behaviors. As established in social science, these behaviors, which include expressions, head motions, and gaze, are not merely decorative. They play a fundamental role in regulating turn-taking, signaling engagement and comprehension, and conveying interpersonal attitudes \cite{Goodwin_1980,hall2019nonverbal,argyle1965eye}. Natural conversation is characterized by a continuous, reciprocal exchange of these subtle cues, such as mutual gaze during listening or a timely nod in response to a key point, as illustrated in Fig.~\ref{fig:cover}. Such cues together form a coordinated and cohesive whole.

Despite the importance of these behaviors, current dialogue head generation methods do not model the semantics of the non-verbal behavior that passes between the two participants. By semantics we mean the social function that a behavior serves in the interaction. Specifically, current dialogue head generation tasks fall into two categories. The first is conditional response generation \cite{peng2025dualtalk,peng2026dyadit}, which takes the behavior signal of one participant as input to synthesize the non-verbal behavior of the other. It therefore captures one side's reaction rather than a bidirectional exchange, and it requires the partner's behavior to be known in advance. The second is joint full-sequence generation \cite{zhou2025interactive,tran2024dim,shan2026talking,zhu2025infp,chu2026unils,teotia2026steer}, which predicts the behavior of both sides from the two audio streams alone. This second category, however, treats the problem as motion synthesis. It optimizes how realistic the generated head animation looks, but the social acts that carry the interaction are never represented, and the coordination between the two generated tracks is never an objective.

To address these limitations, we present DiaHeads, a framework for jointly synthesizing the 3D head animations of both participants from their audio streams. DiaHeads directly generates the low-level continuous parameters defining 3D head pose, expression, and gaze direction for both individuals, and it concatenates the two audio streams and decodes both participants' parameters from one shared latent variable and one shared conditioning vector. A key insight of our work is that incorporating discrete non-verbal cues as auxiliary learning signals significantly guides and enhances the coordination and realism of the synthesized low-level motion. These cues, which include gaze-on, head shaking, nodding, and smiling, are strong semantic anchors for social intent. We represent these cues as four binary variables and fuse them with the continuous parameters in a shared latent space through a product-of-experts posterior, so that the cues carrying the interaction semantics constrain how the continuous motion is generated. The generator additionally replaces the standard Gaussian prior with a flow-based prior over the shared latent variable, which models a temporally dependent and expressive latent distribution rather than an independent Gaussian. The model therefore learns the latent interaction patterns that couple the participants and produces coordinated behaviors such as responsive gaze shifts and contextually appropriate facial expressions.

We target offline, content-generation scenarios, such as pre-visualization for animation or generating dialogues for non-player characters (NPCs) from pre-recorded audio. This distinguishes our work from research on real-time interactive avatars, and allows us to focus on maximizing the perceptual quality and coordinated richness of the generated content.

In summary, this work makes four contributions. First, we propose DiaHeads, a generation model dedicated to the non-verbal behavior of two-person dialogue. It generates the head and gaze parameters of both participants in a single forward pass from the two audio streams, and it uses discrete non-verbal cues as auxiliary signals to improve motion quality and interactional coordination. Second, we propose CANDOR-FG, which extends CANDOR with synchronized FLAME parameters and gaze directions for two-person dialogues. We report its construction, its quality-filtering criteria, and an analysis of the reliability of its estimated targets. Third, beyond reconstruction and audio--lip synchronization, our evaluation protocol reports a mutual-synchrony statistic, a dispersion check, and a reaction-lag statistic. Fourth, comparison experiments, ablation studies, and a user study show that DiaHeads achieves the best performance in generation coordination.

\section{Related Work}
\label{sec:related}

\subsection{Audio-Driven Single-Sided Head Animation}

\noindent\textbf{Speaking head animation.}\quad Early audio-driven facial animation methods primarily focused on lip-sync. VOCA \cite{cudeiro2019capture} used temporal convolutional neural networks (CNNs) to generate coherent animations from speech and a neutral mesh. FaceFormer \cite{fan2022faceformer} employed transformers to model long-range dependencies for expressive sequences, while CodeTalker \cite{xing2023codetalker} incorporated a vector-quantized variational autoencoder (VQ-VAE) to learn discrete motion priors and mitigate over-smoothing. Though achieving high-quality lip-sync, these methods pay limited attention to other facial regions. Other recent efforts include SadTalker \cite{zhang2023sadtalker}, which generates stylized 3D talking-face animation from a single portrait by learning realistic motion coefficients, and SyncTalk \cite{peng2024synctalk}, which emphasizes precise synchronization between lip movements and audio.

\noindent\textbf{Listening head animation.}\quad For listener behavior synthesis, Responsive Listening Head Generation (RLHG) \cite{zhou2022responsive} learned general motion patterns to react to speaker audio. Follow-up works improved output quality via regularization and rendering techniques, as in Perceptual Conversational Head (PCH) \cite{huang2022perceptual}, enhanced diversity using discrete representations, as in Learning to Listen (L2L) \cite{ng2022learning}, or enabled controllable responses with predefined attitudes, as in Emotional Listener Portrait (ELP) \cite{song2023emotional}. Realtalk \cite{geng2023affective} integrated language models and vision-language retrieval, and Ng et al. \cite{ng2023can} adapted large language models for listener motion generation.

While speaking-head methods excel in lip-sync and expression, they overlook gaze, head motion, and listener feedback, whereas listening-head methods prioritize non-verbal cues but treat the partner's behavior only as a one-directional conditioning signal. Neither family captures the mutually conditioned social signals of a real dialogue.

\subsection{Audio-Driven Dialogue Head Animation}

Research on dialogue-level head animation generation falls into two tasks with different problem settings, conditional response generation and joint full-sequence generation.

The \emph{conditional response generation} task focuses on synthesizing the non-verbal behavior of one participant A. Its input includes the dual audio streams and a behavior signal of the other participant B as necessary conditioning. DualTalk \cite{peng2025dualtalk} requires B's ground-truth head motion as input and models the non-verbal behavior of A through cross-modal cross-attention. DyaDiT \cite{peng2026dyadit} instead takes B's pose together with further control signals to guide the generation of A's non-verbal behavior. This task targets real-time interactive avatar scenarios, where one participant is a real user whose behavior is known and only the virtual agent has to be generated. It therefore cannot serve offline pre-scripted content creation, where neither participant's motion is available in advance, and it cannot separate a genuine response to the partner from a plausible listener behavior produced on its own.

The \emph{joint full-sequence generation} task, in contrast, synthesizes complete 3D head behavior sequences for both participants from the two audio streams alone, without any ground-truth behavior sequence as conditional guidance. It matches offline content creation such as animated-film pre-visualization and game NPC dialogue, where creators hold only pre-recorded dual-role audio and need coordinated, full-length animations for both characters in one pass. Within this task, two paradigms can be distinguished according to how they handle roles.

\subsubsection{Speaker--Listener}

In this paradigm, the two participants are assigned fixed roles, a speaker and a listener, and each role is generated by different condition. Zhou et al. \cite{zhou2025interactive} connect an independent speaker generator and an independent listener generator with a role-switching module, at the cost of uncoordinated transitions between turns. DIM \cite{tran2024dim} jointly models speaker and listener actions through masked contrastive learning, but still relies on manual role assignment in dyadic dialogues, which adapts poorly when the two participants speak simultaneously. Talking Together \cite{shan2026talking} controls role change more smoothly, by linearly interpolating two learned speaker and listener embeddings at each time step according to a speaker probability, but it still cannot model the continuous bidirectional coupling of non-verbal behavior in a real dialogue.

\subsubsection{Interlocutor--Interlocutor}
\label{sec:interlocutor}

This paradigm does not distinguish between the speaker and the listener. It always uses the same method to generate behaviors for both interlocutors. INFP \cite{zhu2025infp} predicts the head behavior of both participants through a low-dimensional motion latent space. UniLS \cite{chu2026unils} instead first learns an audio-free autoregressive motion prior and then finetunes it with dual-track audio to generate the head behavior of both sides. Both methods, however, leave the non-verbal interaction semantics unmodeled. They predict motion, but they do not represent the discrete social acts that carry the interaction, and neither evaluates the interactional coordination of the dialogue heads it generates. STEER \cite{teotia2026steer} is the first to bring non-verbal cue labels into this task, factorizing dyadic head behavior into explicit channels for gaze, head rhythm, and emotion, so that a user can steer how an avatar listens, reacts, and engages. However, its labels are given as explicit control signals rather than predicted from audio. The interactional behavior is specified by the user instead of being inferred from the conversation.

Across both paradigms, the interaction semantics of the conversation are never represented. The social behaviors regulate turn-taking and signal engagement, omitting them leaves the generator with no signal for when and how the two heads should respond to each other. DiaHeads is designed specifically for coordinated dyadic dialogue. It uses the non-verbal cue labels of both interlocutors as interactive semantic guidance to model the interactive behaviors of the two sides, which effectively improves the interaction coordination between the two generated dialogue heads.

\subsection{Interactional Coordination and Its Measurement}

Beyond the realism of individual motion, the behavioral coordination between interlocutors is equally critical to the realism of a conversation. Dialogue head generation methods are evaluated with per-participant metrics, such as distribution distances or mean squared errors that measure the realism of each generated head. Each such metric scores one participant against its own target and is therefore blind to the other track. Coordination between the two tracks is what turns individual motion into an exchange. Social psychology shows that interpersonal coordination in non-verbal behavior correlates with rapport, closeness, and trust \cite{fujiwara2026coordination,hendrikse2023becoming,meynard2021predicting,tickle1990nature}. Gaze is a primary regulator of affiliation and turn exchange \cite{argyle1965eye}, and nodding functions as a conventional back-channel signal \cite{Aburumman_2022_08}. To measure such coordination, which is expressed through time-lagged responses, we adopt Dynamic Time Warping (DTW) as the primary coordination metric. DTW has been validated against observer ratings of interpersonal synchrony, and its path features have been shown to predict social outcomes such as trust in economic games \cite{novotny2022identifying,meynard2021predicting}. Considering that DTW can be minimized by a fully mirrored degenerate solution, we add three auxiliary diagnostics that guard against such collapse, namely the Dynamic Mutual Synchrony Score (DMSS), the Mean Absolute Difference (MAD), and the Lag, as described in Section~\ref{sec:metrics}.

\section{The CANDOR-FG Dataset}
\label{sec:dataset}

Effective modeling of dyadic non-verbal behaviors requires datasets with detailed 3D facial parameters and gaze, which are often absent from large-scale conversational corpora. To address this, we introduce CANDOR-FG, a new resource derived from the CANDOR dataset \cite{reece2023candor}. CANDOR-FG provides synchronized FLAME parameters and continuous 3D gaze directions for both participants in a dialogue, which we use to synthesize 3D dialogue heads. It also inherits CANDOR's original crowd-sourced frame-level annotation of conversational behaviors, which we use as the discrete cue layer, so the discrete supervision is human-annotated rather than estimated by our own pipeline.

\subsection{Dataset Construction}

We processed the original long-form CANDOR dialogues to create a dataset suitable for joint behavior generation. Dialogues were segmented into clips of 28--32\,s. The key criterion was that each clip must contain at least a complete ``ABAB'' turn-taking structure based on the provided transcripts. Each segment therefore contains a bidirectional interaction rather than a monologue. This process initially yielded 16{,}852 clips.

We then applied a two-stage quality filter. First, clips containing significant facial occlusion that affects more than 60 consecutive frames, or major camera-view changes, were removed. Second, we rejected clips whose automatic 3D reconstruction was unreliable. A clip was discarded if it contained more than 24 frames on which gaze-direction estimation failed, or more than 60 frames on which FLAME parameter extraction failed. Short gaps within the retained clips were filled by linear interpolation from adjacent frames. The final curated collection contains 14{,}204 high-quality dialogue clips.

To obtain 3D facial geometry we fit the FLAME model to each video frame using the state-of-the-art 3D face reconstruction method SMIRK \cite{retsinas2024smirk}. The process extracts the full set of per-frame parameters. These are shape $\beta\in\mathbb{R}^{300}$, expression $\psi_{expr}\in\mathbb{R}^{50}$, head pose $\theta_{pose}\in\mathbb{R}^{3}$, eyelid $\psi_{eye}\in\mathbb{R}^{2}$, jaw $\theta_{jaw}\in\mathbb{R}^{3}$, and camera $\theta_{cam}\in\mathbb{R}^{3}$. For our head animation task we utilize the subset $f=[\psi_{expr},\theta_{pose},\psi_{eye},\theta_{jaw}]\in\mathbb{R}^{58}$. Because FLAME does not model gaze explicitly, we augmented the data with 3D gaze directions, represented as a pair of angles, pitch ($g_p$) and yaw ($g_y$), estimated for each frame using an L2CS-Net-based model from the MobileGaze framework \cite{valikhujaev2024mobilegaze}. This estimator achieves a mean absolute error of $11.33^{\circ}$ on Gaze360. We applied linear interpolation to fill occasional estimation failures, so each participant has a continuous gaze trajectory $g=(g_p,g_y)$.

\subsection{Annotation Reliability}
\label{sec:dataquality}

Because the continuous targets are produced by automatic estimators rather than motion capture, we quantified their reliability. The discrete non-verbal cue labels do not come from our estimators at all. They originate from CANDOR's original crowd-sourced frame-level annotations. For gaze, two annotators independently inspected 150 sampled frames each, giving 300 verification samples in total. Of these, 89\% were judged correct, 6\% approximately correct, and 5\% incorrect. A comparable visual verification of the FLAME parameters could not be completed, because individual expression, pose, eyelid, and jaw coefficients cannot be judged reliably from a single frame. The rendered mesh is a nonlinear function of the whole parameter vector, so an error in one coefficient can be compensated by another. We therefore rely on SMIRK \cite{retsinas2024smirk} for this component, and report FLAME-based metrics as relative comparisons under one fixed reconstruction pipeline rather than as absolute ground-truth errors.

\subsection{Statistics and Splits}

The CANDOR-FG dataset comprises 14{,}204 dialogue clips. To rigorously evaluate the generalization of the model to unseen individuals we adopt a subject-independent split. By session, the training, validation, and test sets contain 1{,}003, 78, and 77 sessions, i.e., a ratio of about 13:1:1 (86.6\%, 6.7\%, and 6.7\%). By clip, they contain 12{,}386, 909, and 909 clips, i.e., a ratio of about 13.6:1:1 (87.2\%, 6.4\%, and 6.4\%). Hyper-parameters and stage-transition thresholds were tuned on the validation set only. No tuning was performed on the test set. The split indices will be released together with the dataset.

\begin{figure*}[!t]
    \centering
    \includegraphics[width=0.9\textwidth]{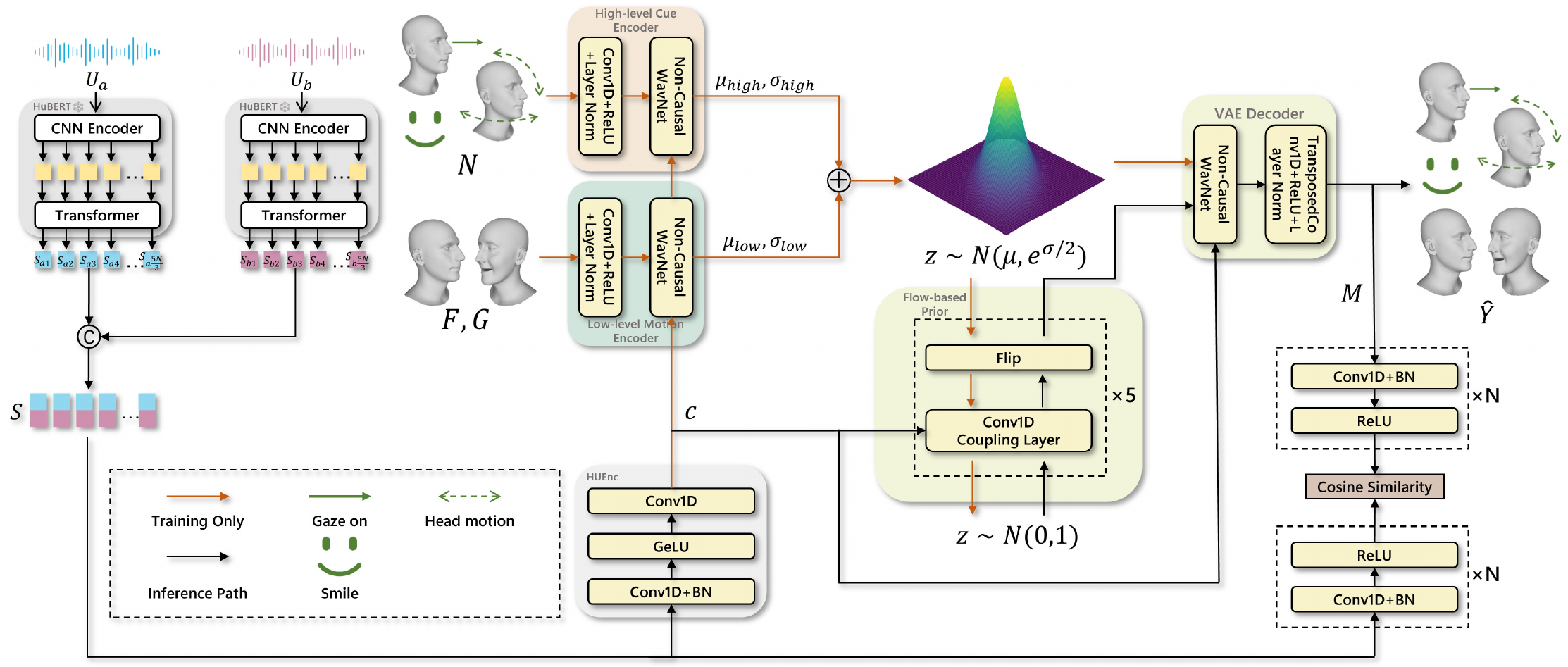}
\caption{Overview of the proposed DiaHeads. The audio streams of both interlocutors are encoded by two frozen HuBERT backbones and fused into a conditioning vector $c$ by a transformer. A dual-branch variational encoder processes the continuous parameters $(F,G)$ and the discrete cues $N$ in separate branches, each conditioned on $c$ through FiLM modulation, and the two branches are combined through a product-of-experts posterior $z\sim\mathcal{N}(\mu,e^{\sigma/2})$, shown as the ``Low-level Motion Encoder'' and ``High-level Cue Encoder''. A flow-based prior built from five residual coupling blocks, each consisting of a flip operation and a Conv1D coupling layer, replaces the standard Gaussian prior, and a shared VAE decoder, conditioned on $c$ through FiLM, jointly reconstructs the head parameters $\hat{Y}$ and the discrete cues of both interlocutors. The units at the bottom right detail the two SyncNet encoders, which are built from 1D convolution, batch normalization, and ReLU.}
    \label{fig:framework}
\end{figure*}

\section{Method}
\label{sec:method}

Our goal is to generate a sequence of low-level continuous parameters, FLAME and gaze direction, for both participants, conditioned on their audio streams. A key aspect of our approach is the incorporation of discrete non-verbal cues as auxiliary learning signals. To this end, our method employs a parameterized FLAME model as the head representation, a pre-trained HuBERT encoder \cite{Hsu_Bolte_Tsai_Lakhotia_Salakhutdinov_Mohamed_2021} to extract speech features, and pitch/yaw angles for gaze. As illustrated in Fig.~\ref{fig:framework}, DiaHeads encodes and concatenates the two audio streams, and it outputs all continuous parameters and auxiliary discrete cues for both participants in a single forward pass. Its four components each contribute to modeling the interaction rather than only to per-participant realism. The generator decodes both participants from one shared latent variable and one conditioning vector built from both audio streams (Section~\ref{sec:generator}). The four discrete non-verbal cues are fused with the continuous parameters through a product-of-experts posterior (Section~\ref{sec:poe}), and a flow-based prior replaces the Gaussian prior over that latent variable (Section~\ref{sec:flow}). A pre-trained SyncNet then adds an audio--lip constraint on each participant separately (Section~\ref{sec:audio-lip}).

\subsection{Problem Formulation}
\label{sec:formulation}

Let $U_a^{1:\frac{5}{3}N}=\{u^1_a,\dots,u^{\frac{5}{3}N}_a\}$ and $U_b^{1:\frac{5}{3}N}=\{u^1_b,\dots,u^{\frac{5}{3}N}_b\}$ denote the audio sequences of participants A and B, where $N$ is the video frame count, the audio frame count is $\frac{5}{3}N$, and each $u\in\mathbb{R}^{D}$ with $D$ the audio sample rate. Our model then extracts speech features $S_a^{1:\frac{5}{3}N}=\{s^1_a,\dots,s^{\frac{5}{3}N}_a\}$ and $S_b^{1:\frac{5}{3}N}=\{s^1_b,\dots,s^{\frac{5}{3}N}_b\}$ from $U_a$ and $U_b$, where $s\in\mathbb{R}^{1024}$. Subsequently, $S_a$ and $S_b$ are concatenated to generate the output sequence $Y^{1:N}=[Y_a^{1:N},Y_b^{1:N}]=\{y^1_a,\dots,y^{N}_a,y^1_b,\dots,y^{N}_b\}$, where $y=[f,g,n]$, $f\in\mathbb{R}^{58}$, $g\in\mathbb{R}^2$, and $n\in\{0,1\}^4$. Here $f$ denotes the FLAME parameters, which include expression, pose, eyelid, and jaw. $g$ represents gaze direction, namely pitch and yaw angles. $n$ encodes non-verbal cues, namely gaze-on, head shaking, nodding, and smiling. These cues provide auxiliary supervision during training and are predicted as well. For brevity, we refer to $Y$ as the \emph{joint head parameters} and to $Y_a$/$Y_b$ as the \emph{head parameters} throughout the paper.

\subsection{Joint Head Parameter Generator}
\label{sec:generator}

We propose DiaHeads, a variational autoencoder based joint head parameter generator that maps the dual-speaker audio inputs $U=[U_a,U_b]$ to the joint head parameters $Y$. DiaHeads treats the dual-speaker audio features $S=[S_a,S_b]$ as conditional signals that drive a shared latent space, from which the joint head parameters $Y$ are jointly decoded. First, an audio encoder maps the two HuBERT feature streams $S$ into a conditioning vector $c$.
\begin{equation}
{c} = \text{HuEnc}(S_a, S_b).
\end{equation}
This conditioning vector $c$, together with the ground-truth joint head parameters $Y$, is then processed by our dual-branch VAE encoder, which maps them into the latent space. The encoder employs two parallel branches that reflect the hierarchical nature of non-verbal communication, a low-level motion encoder for the continuous facial parameters $F$ and gaze direction $G$, and a high-level cue encoder for the discrete non-verbal cues $N$. Both branches are conditioned on $c$ and use the same frame-wise validity mask $\mathbf{m}$. Formally, the low-level motion encoder outputs
\begin{equation}
{\mu}_{\text{low}}, {\sigma}_{\text{low}} = \text{Enc}_{\text{low}}(F, G, \mathbf{m}, {c}),
\end{equation}
and the high-level cue encoder outputs
\begin{equation}
{\mu}_{\text{high}}, {\sigma}_{\text{high}} = \text{Enc}_{\text{high}}(N, \mathbf{m}, {c}).
\end{equation}
Here $\sigma$ denotes log-variance. To obtain an identifiable latent representation, we truncate the reparameterized sample with a hyperbolic tangent squashing function, as in standard VAEs. Because both tracks are decoded from the same latent variable, and the conditioning vector is built from both audio streams, the joint model can represent dependence between the two participants that two independently trained per-participant models cannot.

\subsubsection{Fusing the Two Posterior Branches}
\label{sec:poe}

The two branches describe the same latent variable from complementary evidence, so they are combined as a product of experts rather than by simple averaging. Multiplying the two diagonal Gaussians $\mathcal{N}(\mu_{\text{low}},e^{\sigma_{\text{low}}})$ and $\mathcal{N}(\mu_{\text{high}},e^{\sigma_{\text{high}}})$ and normalizing yields a Gaussian whose natural parameters are the sums of the individual natural parameters, i.e.
\begin{equation}
\begin{split}
    \mu &= \frac{\mu_{\text{low}}/e^{\sigma_{\text{low}}}+\mu_{\text{high}}/e^{\sigma_{\text{high}}}}{1/e^{\sigma_{\text{low}}}+1/e^{\sigma_{\text{high}}}}, \\
    \sigma &= -\log\!\left(\frac{1}{e^{\sigma_{\text{low}}}}+\frac{1}{e^{\sigma_{\text{high}}}}\right).
\end{split}
    \label{eq:poe}
\end{equation}
Equation~\eqref{eq:poe} performs a weighted sum of the means and a harmonic mean of the log-variances, which is a nonlinear product of experts (PoE). Each encoder is therefore an independent expert. The low-level encoder defines the audio-to-motion mapping, while the high-level encoder constrains which latent subspaces are plausible, for instance the smiling region. Because $N$ collects the cues of both participants, the cue expert constrains the latent by the pair of social states rather than by each participant in isolation. The cue branch therefore shapes the latent according to the exchange between the two participants, which is where the interaction semantics enter the generation. Only regions that both experts find probable are reinforced. This imposes a distribution-level constraint instead of a per-frame vector addition, which is what allows the discrete cue branch to shape the latent geometry without over-riding the motion evidence.

The latent variable $z$ is sampled by reparameterization.
\begin{equation}
z = \mu + e^{\sigma/2} \odot \epsilon, \quad \epsilon \sim \mathcal{N}(0, I).
\end{equation}
Here ${\mu}$ and ${\sigma}$ are the mean and log-variance of the approximate posterior distribution $q(z \mid Y, c)$. The VAE decoder reconstructs the joint head parameters $\hat Y$ conditioned on the latent variable and audio context.
\begin{equation}
    \hat{Y} = \text{Dec}({z}, {c}, \mathbf{m}).
\end{equation}

\subsubsection{Flow-Based Prior}
\label{sec:flow}

To address the temporal inconsistency and limited diversity caused by the independent Gaussian prior in standard VAEs, we adopt a flow-based prior module conditioned on the audio features $c$ to learn a temporally dependent, expressive prior distribution $p({z}\mid{c})$. Our target output consists of low-dimensional sequential parameters including FLAME and gaze data, unlike high-dimensional data such as images or video. Flow models suit this setting because they provide exact latent density estimation and efficient sampling, learning smooth and temporally coherent human motion manifolds without the iterative denoising steps typical of diffusion models. Because the prior is learned over the shared latent variable, it models the temporal structure of the coupled motion rather than treating frames as independent. Furthermore, to ensure precise temporal alignment between the generated lip movements and the input speech, we integrate a pre-trained SyncNet module, whose details are provided in Section~\ref{sec:audio-lip}.

\subsection{Audio--Lip Motion SyncNet}
\label{sec:audio-lip}

We implement a SyncNet to measure the temporal alignment between audio and lip movements, following the SyncExpert architecture \cite{prajwal2020lip}. Our SyncNet utilizes the jaw and expression parameters from FLAME as the lip motion proxy, denoted as $M=\{\psi_{expr},\theta_{jaw}\}\in\mathbb{R}^{53}$. The SyncNet architecture comprises two parallel encoders, a lip-motion encoder processing FLAME parameters and an audio encoder handling HuBERT features. Both encoders employ stacked 1D convolutional layers followed by batch normalization and ReLU activations. A 6-frame lip proxy sequence $M^{1:6}$ and its corresponding 10-frame HuBERT feature clip $S^{1:10}$ are input to the encoders, generating embeddings $e_m\in\mathbb{R}^{512}$ and $e_s\in\mathbb{R}^{512}$. The synchronization score is computed via cosine similarity between the embeddings and then mapped to the unit interval.
\begin{equation}
d = \text{cosine\_similarity}(e_m,e_s) = \frac{e_m \cdot e_s}{\lVert e_m\rVert \times \lVert e_s\rVert}.
\end{equation}
The training loss is defined as binary cross-entropy over the normalized similarity score.
\begin{equation}
    \mathcal{L}_\text{syncNet} = \text{BCE}(d, l),
\end{equation}
where $l\in\{0,1\}$ denotes the binary synchronization label.

\subsection{Training Process}

We employ a multi-stage training strategy. In the pre-training phase, the SyncNet module is first trained independently, so that it can accurately assess the synchronization between facial motion and speech. Subsequently, the main DiaHeads model is trained with the synchronization loss weight initially set to zero, so that the model first learns basic reconstruction. In the final joint optimization phase, triggered when the validation sync loss drops below a predefined threshold, the synchronization loss term is activated and the model is trained end to end over all objectives. The total loss is a weighted sum of multiple objectives.
\begin{equation}
\begin{split}
       \mathcal{L}_{\text{total}} = &\lambda_{\text{kl}} \mathcal{L}_{\text{kl}}
+ \lambda_{\text{mse}}\mathcal{L}_{\text{mse}}
       +\lambda_{\text{cont}}\mathcal{L}_{\text{cont}} \\
       &+ \lambda_{\text{bce}}\mathcal{L}_{\text{bce}}
+ \lambda_{\text{sync}} \mathcal{L}_{\text{sync}},
\end{split}
\end{equation}
The KL divergence loss measures the discrepancy between the posterior distribution $q({z}\mid{Y},{c})$ and the flow-based prior $p({z}\mid{c})$.
\begin{equation}
    \mathcal{L}_{\text{kl}} = \mathbb{E}_{q}\!\left[\log \frac{q({z}\mid{Y},{c})}{p({z}\mid{c})}\right].
\end{equation}
The mean squared error loss measures the reconstruction error.
\begin{equation}
    \mathcal{L}_{\text{mse}} = \sum_{i,j} \mathbf{m}_{i,j}\cdot
    \lVert\hat{{\{f,g\}}}_{i,j}-{\{f,g\}}_{i,j}\rVert^2 .
\end{equation}
The continuity loss encourages temporal smoothness.
\begin{equation}
\begin{split}
    \mathcal{L}_{\text{cont}} =  \sum_{i,j>0} \mathbf{m}_{i,j}\cdot &\lVert\Delta\hat{{y}}_{i,j}
- \Delta{y}_{i,j}\rVert^2 + \sum_i \lVert\hat{{y}}_{i,0}-{y}_{i,0}\rVert^2,
\end{split}
\end{equation}
where $\Delta{y}_{i,j}={y}_{i,j}-{y}_{i,j-1}$ denotes the first-order temporal difference. The BCE loss targets the discrete non-verbal cue indicators $n$.
\begin{equation}
\begin{split}
    \mathcal{L}_{\text{bce}} = - \sum_{i,j,k} &\mathbf{m}_{i,j}
    \big[\, n_{i,j,k} \log \text{sigmoid}(\hat{n}_{i,j,k}) \\
    &+(1-n_{i,j,k}) \log(1-\text{sigmoid}(\hat{n}_{i,j,k})) \,\big],
\end{split}
\end{equation}
where the sigmoid function is applied element-wise. Finally, the synchronization loss evaluates the alignment between generated mouth movements and speech using the pre-trained SyncNet, where $\text{SyncNet}(\cdot)$ returns the synchronization score.
\begin{equation}
\begin{split}
    \mathcal{L}_{\text{sync}} = &\ \text{BCE}(\text{SyncNet}(\hat{M}_a,{S_a}), 1) \\
    &+ \text{BCE}(\text{SyncNet}(\hat{M}_b,{S_b}), 1).
\end{split}
\end{equation}

Every term above is defined on the joint sequence $Y=[Y_a,Y_b]$, so the objective fits a single distribution over the two tracks rather than two per-participant models.

\section{Experiments}
\label{sec:experiments}

\subsection{Experimental Setup}

Experiments are conducted on the CANDOR-FG dataset. The proposed DiaHeads framework is compared against state-of-the-art methods, DIM \cite{tran2024dim}, SadTalker \cite{zhang2023sadtalker}, and SyncTalk \cite{peng2024synctalk}. To ensure a fair comparison, all baselines are retrained on CANDOR-FG. SadTalker and SyncTalk are single-person methods and process each speaker independently through two separate inference passes, whereas DIM is the closest method in the joint-sequence line. We restrict the comparison to baselines whose training code is publicly available, so that every number in Table~\ref{tab:quantitative} is produced under our data pipeline. INFP \cite{zhu2025infp}, the method of Zhou et al. \cite{zhou2025interactive}, and DualTalk \cite{peng2025dualtalk} do not release training code, and DualTalk additionally requires the partner's ground-truth head motion, which is unavailable under our task setting. A fair retraining comparison with them is therefore not possible. Given that the proposed framework outputs joint head parameters, the same data processing pipeline is applied to extract corresponding head parameters from the outputs of the baseline methods for metric evaluation.

\begin{table}[!t]
\centering
\caption{Quantitative Comparison of State-of-the-Art Methods}
\label{tab:quantitative}
\small
\setlength{\tabcolsep}{4.5pt}
\begin{tabular}{lcccc}
\toprule
\textbf{Model} & \makecell{\textbf{FLAME}\\\textbf{MSE}$\downarrow$} &
\makecell{\textbf{Gaze}\\\textbf{MSE}$\downarrow$} &
\makecell{\textbf{Sync}\\\textbf{Error}$\downarrow$} &
\makecell{\textbf{Interaction}\\\textbf{DTW}$\downarrow$} \\
\midrule
\multicolumn{5}{l}{\footnotesize \textit{Single-speaker talking heads}} \\
SadTalker & \textbf{0.5523} & 0.1047 & 1.3352 & 0.6963 \\
SyncTalk  & 0.7603 & 0.0857 & 1.5703 & 0.8717 \\
\midrule
\multicolumn{5}{l}{\footnotesize \textit{Dialogue heads}} \\
DIM      & 0.8164 & 0.0858 & 1.2769 & 0.7526 \\
DiaHeads (Ours) & 0.6662 & \textbf{0.0808} & \textbf{0.7995} & \textbf{0.3171} \\
\bottomrule
\end{tabular}
\end{table}

\subsection{Evaluation Metrics}
\label{sec:metrics}

We evaluate the 3D dialogue head animation model with two groups of metrics. \emph{Fidelity and synchronization} are per-participant. The FLAME MSE measures the fidelity of facial geometry and pose reconstruction by calculating the mean squared error between predicted and ground-truth FLAME parameters, averaged across both interlocutors. The Gaze MSE assesses the accuracy of generated eye movements via the mean squared error of predicted gaze directions, averaged over both participants. The Sync Error quantifies the temporal alignment between generated lip motions and the input audio using the SyncNet model.

\emph{Coordination} metrics capture the relation between the two participants. The primary one, Interaction DTW, quantifies the behavioral coordination between interlocutors by computing the Dynamic Time Warping distance between their generated motion sequences, with FLAME and gaze weighted at 0.7 and 0.3. The DTW distance is particularly suited for measuring behavioral coordination as it accommodates natural, nonlinear time lags in reactive behaviors \cite{novotny2022identifying}. This choice is grounded in social psychology, where DTW-based synchrony has been shown to predict social outcomes such as trust \cite{meynard2021predicting}. A lower DTW indicates stronger coupling and a more cohesive dyad.

Because a warping-based distance can in principle be reduced by degenerate solutions, we also report three auxiliary diagnostics computed on the same generated sequences. The \emph{Dynamic Mutual Synchrony Score} (DMSS) is the maximum lagged cross-correlation between the two participants' motion traces over a lag window of $\tau\in[-30,30]$ frames. A higher value means stronger coordination, and unlike DTW it is insensitive to a constant offset and directly exposes the temporal structure of the coupling. The \emph{Mean Absolute Difference} (MAD) is the mean absolute frame-wise difference between the two participants' motion traces. It approaches zero if the generator collapses both participants onto the same motion, so a healthy model should stay within the natural range observed in real dialogues, which we measure to be roughly 0.3--0.6. The \emph{Lag} is the time delay at which the maximum cross-correlation is achieved. A very short lag indicates mechanical mirroring, while a very long lag indicates that the two tracks are effectively disconnected.

\subsection{Quantitative Results}

Table~\ref{tab:quantitative} presents the quantitative comparison between our DiaHeads and state-of-the-art methods, covering single-speaker talking-head models and dyadic dialogue head animation models. The results show clear performance gaps across methods, which stem from the fundamental differences in their modeling paradigms for conversational scenarios.

For single-speaker talking-head methods, SadTalker achieves the lowest FLAME MSE among all compared models. This result is consistent with its primary design objective, as SadTalker is optimized exclusively for high-fidelity facial animation of a single speaker without modeling dyadic interaction dynamics. However, this design introduces significant limitations within dialogue scenarios. Consequently, the method yields suboptimal performance in both synchronization and interaction metrics. Furthermore, it frequently generates inaccurate lip movements during the listening turn of the character. This weakens both audiovisual consistency and conversational coordination. SyncTalk encounters even more severe challenges in this setting, which illustrates the inherent limitations associated with adapting single-speaker models to dyadic dialogue generation.

For dedicated dyadic dialogue models, DIM achieves a moderate Sync Error, but still yields a high Interaction DTW score. This indicates that its speaker--listener alternation paradigm cannot capture the reciprocal and continuously coupled nature of non-verbal behaviors in human conversations.

In contrast, our DiaHeads achieves the best or competitive performance across all metrics except FLAME MSE. It outperforms all baselines by a large margin in Sync Error. Relative to DIM, the closest dyadic baseline, Sync Error drops from 1.2769 to 0.7995, a reduction of 37.4\%. Most notably, it sets a new state of the art in Interaction DTW, the core metric for evaluating conversational coordination, reducing the distance from 0.7526 (DIM) to 0.3171, by 57.9\%. The lowest Gaze MSE among all methods further validates our model's effectiveness in synthesizing context-aware eye movements.

The slightly higher FLAME MSE of DiaHeads relative to SadTalker stems from the dyadic interaction constraints embedded in our framework. Our model is trained to generate head poses that naturally face the interlocutor, but in a portion of CANDOR-FG clips the ground-truth head poses are in side-view or skewed orientations rather than front-facing. SadTalker, a single-speaker model with no notion of dyadic state, simply aligns with these side-view annotations, whereas DiaHeads prioritizes the social plausibility of the interaction and tends to generate front-facing head poses. This difference in modeling objectives accounts for the higher FLAME MSE, and it also illustrates our core hypothesis that realistic dyadic dialogue animation requires explicit modeling of mutual non-verbal feedback, not only accurate individual motion synthesis.

\begin{figure}[!t]
  \centering
  \includegraphics[width=0.98\columnwidth]{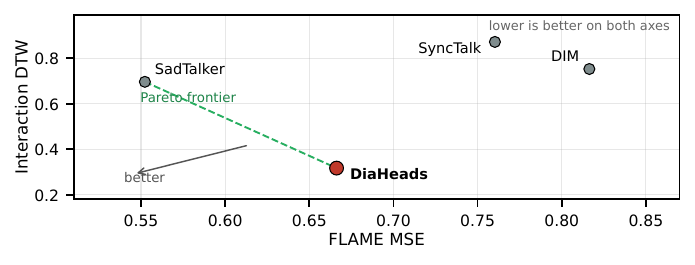}
\caption{Fidelity--coordination trade-off on CANDOR-FG. Each marker is one method, positioned by its FLAME MSE and its Interaction DTW. Only our method and SadTalker lie on the dashed Pareto frontier, and our method achieves the lowest Interaction DTW at a modest cost in FLAME MSE.}
  \label{fig:tradeoff}
\end{figure}

Fig.~\ref{fig:tradeoff} summarizes the same comparison as a fidelity--coordination trade-off. Our method and SadTalker are the only two non-dominated methods, and our method achieves the lowest Interaction DTW at a modest cost in FLAME MSE.

\subsection{Coordination Diagnostics}
\label{sec:coordination}

\begin{figure}[!t]
  \centering
  \includegraphics[width=0.98\columnwidth]{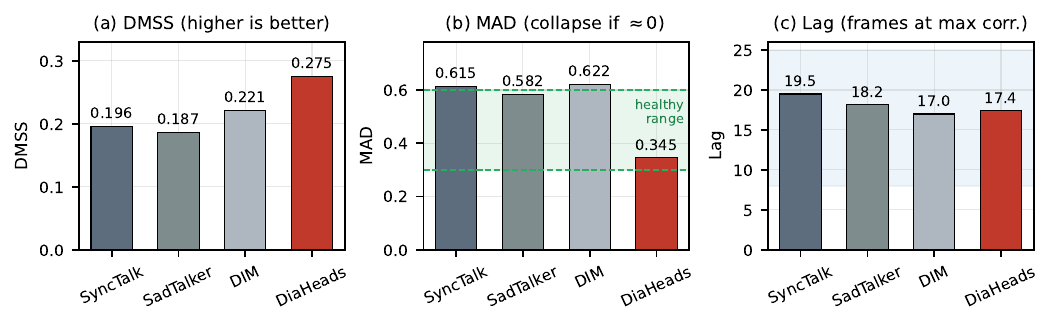}
\caption{Coordination diagnostics. (a) DMSS, the maximum lagged cross-correlation between the two participants over $\tau\in[-30,30]$ frames, where higher is better. (b) MAD, the mean absolute frame-wise difference between the two participants. Values near zero indicate that the generator collapses both participants onto the same motion, and the shaded band marks the natural range in real dialogues. (c) Lag, the delay at which the maximum correlation occurs. Both mechanical mirroring and complete disconnection are undesirable. DiaHeads achieves the strongest lagged coupling without approaching collapse.}
  \label{fig:coord}
\end{figure}

\begin{table}[!t]
\centering
\caption{Coordination Diagnostics and Collapse Check}
\label{tab:coord}
\small
\setlength{\tabcolsep}{4pt}
\begin{tabular}{lccc}
\toprule
\textbf{Method} & \makecell{\textbf{DMSS}$\uparrow$} & \textbf{MAD} & \textbf{Lag} \\
\midrule
DiaHeads (Ours) & \textbf{0.275} & 0.345 & 17.4 \\
DIM      & 0.221 & 0.622 & 17.0 \\
SyncTalk & 0.196 & 0.615 & 19.5 \\
SadTalker & 0.187 & 0.582 & 18.2 \\
\bottomrule
\end{tabular}
\end{table}

Table~\ref{tab:coord} and Fig.~\ref{fig:coord} report the coordination diagnostics. DiaHeads achieves the highest DMSS, exceeding the second-best method by roughly 20\% and the weakest baseline by about a third. This is direct evidence that the two tracks are mutually conditioned rather than merely plausible in isolation.

The MAD column rules out the degenerate explanation for the low DTW. DiaHeads has the lowest MAD, yet it remains inside the natural range of 0.3--0.6. The two participants are not collapsed onto each other, and the synchronization constraint together with the reconstruction loss suppresses such degeneration. The baselines, by contrast, all sit above the upper end of that range, i.e., they preserve inter-participant separation but without coupling it, which is consistent with their weaker DMSS. The Lag column shows that all methods place their maximum correlation in the 17--20 frame range, so no method achieves its coupling through zero-lag copying. DiaHeads lies near the lower end of that range, corresponding to a physiologically plausible reaction delay, and its advantage is therefore not an artifact of mechanical mirroring.

\subsection{Qualitative Results}

\begin{figure}[!t]
    \centering
    \includegraphics[width=0.98\columnwidth]{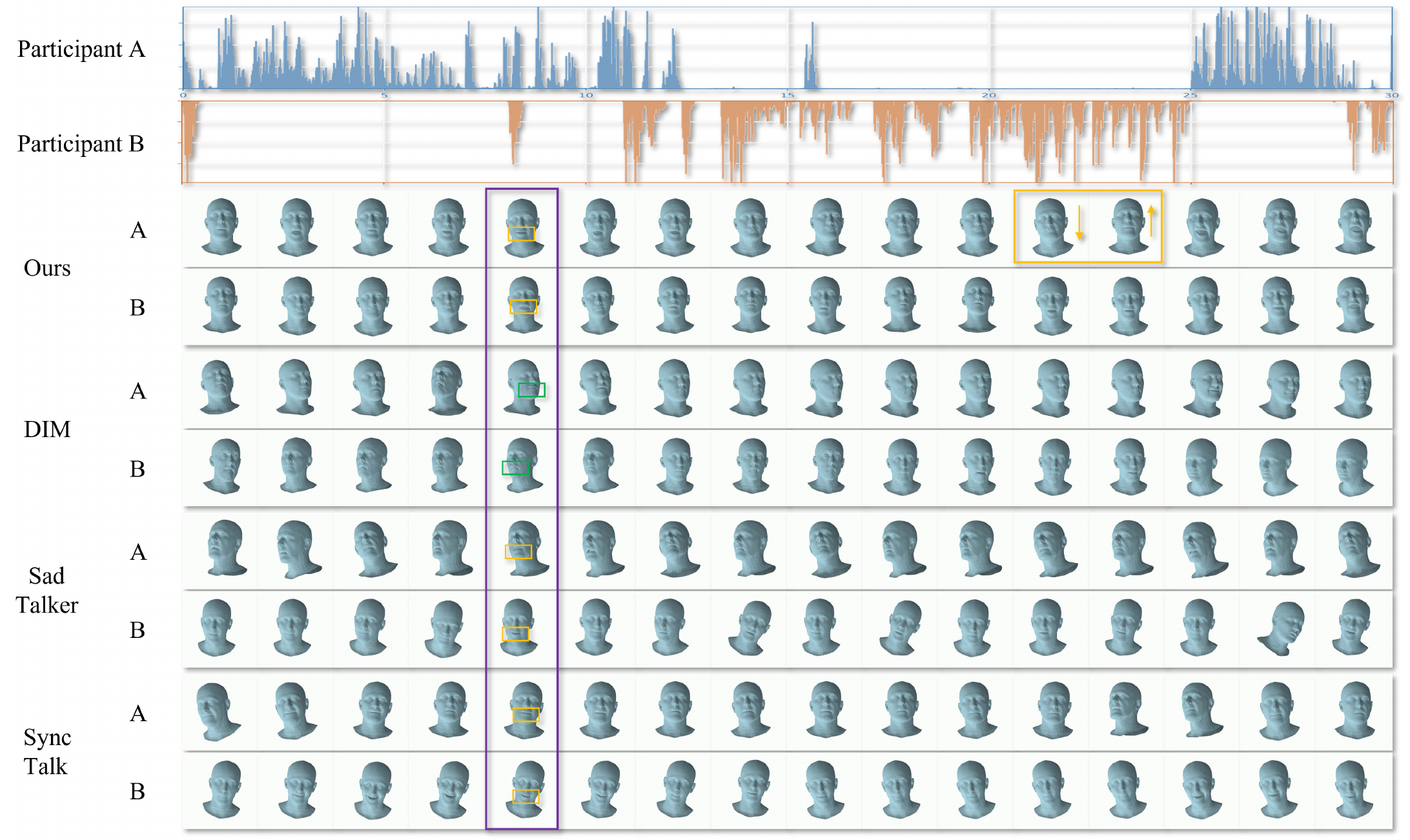}
\caption{Qualitative comparison of generated 3D head animations over a 30\,s dialogue clip. Participant A is shown in blue and participant B in orange. Each method occupies two rows labeled A and B, sampled every 2.5\,s. Our method produces appropriate mouth movements during overlapping speech at 8\,s, marked by the purple box, and responsive nodding with mutual gaze at 20--23\,s, marked by the yellow box, while the baselines show incorrect lip-sync and lack interactive behaviors.}
    \label{fig:Qualitative-Results}
\end{figure}

As illustrated in Fig.~\ref{fig:Qualitative-Results}, our method maintains lip-sync with speech across all dialogue scenarios. A representative case occurs around 8\,s, highlighted by the purple box, where participant B makes a brief interjection while participant A holds the speaking turn from 0 to 12\,s. Our method animates participant B's mouth movements while keeping this participant's gaze on participant A, forming a natural back-channeling response and showing that the model captures the subtle dynamics of conversational engagement.

During the 20--23\,s interval, highlighted by the yellow box, participant A produces a timely nodding response while maintaining mutual gaze with the speaker. Such coordinated behavior is what makes active listening visible. The baselines lack this feedback. Although SadTalker and SyncTalk do open the mouth at the 8\,s interjection, they fail to establish gaze toward the speaker. DIM's speaker--listener paradigm cannot express the subtlety of back-channeling responses. None of them produces context-appropriate nodding or gaze patterns. These observations are consistent with our quantitative results. The superior Interaction DTW and Sync Error indicate that joint generation is what preserves the natural rhythm and reciprocity of conversation.

\subsection{Ablation Study}

\begin{table}[!t]
\centering
\caption{Ablation Study on the Model Components.}
\label{tab:ablation}
\small
\setlength{\tabcolsep}{4.5pt}
\begin{tabular}{lcccc}
\toprule
\textbf{Model} & \makecell{\textbf{FLAME}\\\textbf{MSE}$\downarrow$} &
\makecell{\textbf{Gaze}\\\textbf{MSE}$\downarrow$} &
\makecell{\textbf{Sync}\\\textbf{Error}$\downarrow$} &
\makecell{\textbf{Interaction}\\\textbf{DTW}$\downarrow$} \\
\midrule
Full   & \textbf{0.6662} & \textbf{0.0808} & 0.7995 & \textbf{0.3171} \\
noFlow & 0.8254 & 0.0964 & 0.8041 & 0.5082 \\
noSync & 0.8926 & 0.0968 & 1.2301 & 0.7719 \\
Ind.   & 0.9503 & 0.1064 & 0.8941 & 1.3505 \\
noGaze & 0.8528 & n/a & \textbf{0.7472} & 1.0400 \\
\bottomrule
\end{tabular}
\end{table}

\begin{table}[!t]
\centering
\caption{Ablation on the High-Level Cue Configuration}
\label{tab:cue_ablation}
\small
\setlength{\tabcolsep}{4.5pt}
\begin{tabular}{lcccc}
\toprule
\textbf{Model} & \makecell{\textbf{FLAME}\\\textbf{MSE}$\downarrow$} &
\makecell{\textbf{Gaze}\\\textbf{MSE}$\downarrow$} &
\makecell{\textbf{Sync}\\\textbf{Error}$\downarrow$} &
\makecell{\textbf{Interaction}\\\textbf{DTW}$\downarrow$} \\
\midrule
Full (all cues) & \textbf{0.6662} & 0.0808 & 0.7995 & \textbf{0.3171} \\
Gaze only  & 0.6869 & 0.0797 & 0.8030 & 0.3520 \\
Head only  & 0.6676 & \textbf{0.0789} & \textbf{0.7832} & 0.3509 \\
Smile only & 0.6799 & 0.0790 & 0.8024 & 0.3497 \\
No cues & 0.6782 & 0.0821 & 0.8045 & 0.3675 \\
\bottomrule
\end{tabular}
\end{table}

\label{sec:ablation}

We ablate the four mechanisms on which DiaHeads depends. The variants are \textsf{noFlow}, \textsf{noSync}, \textsf{Ind.}, and \textsf{noGaze}. The \textsf{noFlow} variant removes the flow-based prior and falls back to a standard Gaussian prior. The \textsf{noSync} variant removes the SyncNet loss. The \textsf{Ind.} variant denotes independent generation, which zeroes the partner's audio, runs inference twice, and concatenates each side's own half-prediction, so that no interaction can be modeled. The \textsf{noGaze} variant removes the gaze targets while keeping the FLAME targets.

\begin{figure}[!t]
  \centering
  \includegraphics[width=0.98\columnwidth]{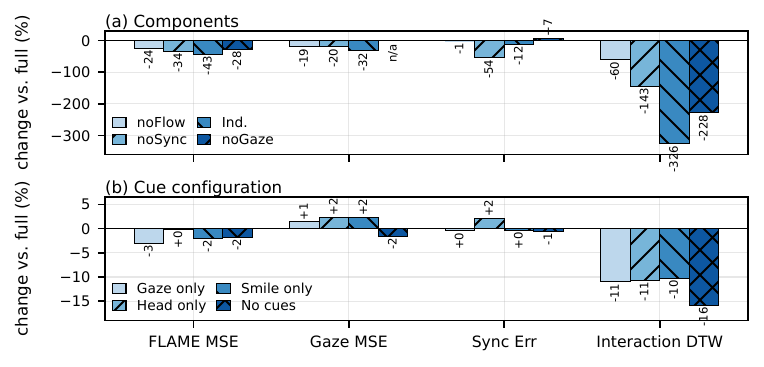}
\caption{Component and cue ablations, as the change in each metric relative to the full model, in percent. Bars below zero are worse and bars above zero are better. (a) Removing one component at a time. (b) Keeping only one cue family at a time.}
  \label{fig:ablations}
\end{figure}

Table~\ref{tab:ablation} and Fig.~\ref{fig:ablations} summarize the results. Removing the flow prior, denoted \textsf{noFlow}, degrades FLAME MSE by 23.9\% and Interaction DTW by 60.3\%, confirming that the flow prior is essential for coordination and not merely a diversity mechanism. Removing the synchronization loss, denoted \textsf{noSync}, increases Sync Error by 53.9\% and Interaction DTW by 143\%, so the audio--lip constraint also carries coordination information. Removing gaze supervision, denoted \textsf{noGaze}, yields the best Sync Error, 6.5\% better than the full model, which is expected because gaze is not an input to the SyncNet term. Its Interaction DTW is computed on FLAME only, since a gaze-free model has no gaze channel to weight, and is therefore not directly comparable to the others. Finally, the \textsf{Ind.} variant, in which each participant is generated from its own audio alone, is worst on every metric, with FLAME MSE up 42.6\% and Interaction DTW up 326\%, the clearest evidence that the observed coordination comes from joint generation rather than from per-participant audio-driven realism.

We also ablate the high-level cue supervision itself, examining four variants, gaze-related cues only, head motion only, smiles only, and no high-level non-verbal cues at all. As summarized in Table~\ref{tab:cue_ablation} and Fig.~\ref{fig:ablations}, the full model achieves the best overall performance, lowering Interaction DTW by approximately 10.0\%, 9.6\%, 9.3\%, and 13.7\% relative to the gaze-only, head-only, smile-only, and no-cue variants, respectively. The head-only variant stays closest to the full model, indicating that head motions such as nodding and shaking play a foundational role in stabilizing facial dynamics and structuring turn-taking behavior. The gaze-only variant achieves competitive gaze accuracy but degrades clearly in interaction synchrony, indicating that gaze alone cannot sustain the reciprocal feedback loop required for natural dialogue. The smile-only variant improves only lip synchronization, an effect due to lip-region articulation during smiling, and does not raise broader interaction quality. The no-cue variant is worst in interaction synchrony despite reasonable low-level facial and gaze motion. Audio can therefore drive basic facial animation, but the absence of structured non-verbal feedback severely weakens the perception of natural interaction, which makes jointly modeling multiple non-verbal cues in a unified context essential for generating socially coherent dialogues.

These ablations affirm that non-verbal cues are essential for realistic dyadic animation, with head motions particularly central, and validate our core argument that realistic dialogue generation requires modeling mutual non-verbal feedback between two participants treated on equal terms.

\subsection{Human Subject Evaluation}
\label{sec:human_evaluation}

To verify the perceptual quality of our generated animations, especially the naturalness of interpersonal coordination, we conduct a human subject evaluation.

\subsubsection{Experiment Setup}

We recruit 16 valid participants (aged 18--35, with normal or corrected vision and hearing), who watch 28 videos in fully randomized order. The set contains 7 test clips for each of the four methods, DiaHeads, DIM, SadTalker, and SyncTalk. All stimuli are anonymized and standardized to the same resolution, frame rate, and audio volume. The evaluation consists of a 5-point Likert Mean Opinion Score (MOS) rating on overall animation naturalness and dyadic interactional coordination, and a forced-choice preference test that selects the most natural and realistic animation among the four method outputs for each clip. The full protocol is provided in the supplementary material. All participants gave verbal informed consent before the session.

\subsubsection{Results and Analysis}

\begin{table}[!t]
\centering
\caption{Results of the Human Subject Evaluation. MOS Values Are Reported as Mean $\pm$ Standard Deviation. The Best Results Are Marked in Bold}
\label{tab:human_evaluation}
\small
\setlength{\tabcolsep}{3.5pt}
\begin{tabular}{lccc}
\toprule
Model & \makecell{Overall\\Naturalness} & \makecell{Interactional\\Coordination} & \makecell{Blind Preference\\Votes} \\
\midrule
SadTalker & $3.10 \pm 1.05$ & $3.17 \pm 1.11$ & 33 \\
SyncTalk  & $2.12 \pm 1.09$ & $2.36 \pm 1.01$ & 4 \\
DIM       & $2.92 \pm 1.13$ & $2.84 \pm 1.12$ & 17 \\
\midrule
DiaHeads (Ours) & $\mathbf{3.73 \pm 1.07}$ & $\mathbf{3.67 \pm 1.11}$ & $\mathbf{58}$ \\
\bottomrule
\end{tabular}
\end{table}

\begin{table}[!t]
\centering
\caption{Pairwise Statistical Comparison of the MOS Ratings Against DiaHeads, Bonferroni-Corrected}
\label{tab:human_stats}
\small
\setlength{\tabcolsep}{3.5pt}
\begin{tabular}{llccc}
\toprule
\textbf{Dimension} & \textbf{Baseline} & \textbf{Cohen's $d$} & \textbf{$t$} & \textbf{$p$} \\
\midrule
\multirow{3}{*}{\makecell{Overall\\Naturalness}}
  & SadTalker & 0.59 & 4.45 & $<10^{-4}$ \\
  & DIM       & 0.74 & 5.51 & $<10^{-5}$ \\
  & SyncTalk  & 1.49 & 11.16 & $<10^{-5}$ \\
\midrule
\multirow{3}{*}{\makecell{Interactional\\Coordination}}
  & SadTalker & 0.45 & 3.37 & $5\times10^{-3}$ \\
  & DIM       & 0.74 & 5.57 & $<10^{-5}$ \\
  & SyncTalk  & 1.23 & 9.24 & $<10^{-5}$ \\
\bottomrule
\end{tabular}
\end{table}

\begin{figure}[!t]
  \centering
  \includegraphics[width=0.98\columnwidth]{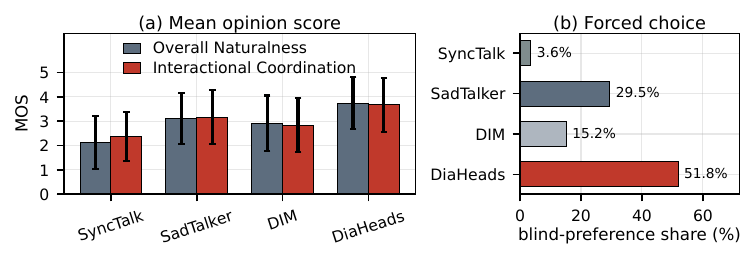}
\caption{Human subject evaluation. (a) Mean opinion score with standard deviation on overall animation naturalness and dyadic interactional coordination. (b) Share of the blind forced-choice votes. Both panels use the same participants and clips as Tables~\ref{tab:human_evaluation} and~\ref{tab:human_stats}.}
  \label{fig:human_eval}
\end{figure}

As shown in Tables~\ref{tab:human_evaluation} and~\ref{tab:human_stats} and Fig.~\ref{fig:human_eval}, DiaHeads reaches a mean score of $3.67\pm1.11$ in \emph{Interactional Coordination}, significantly higher than DIM, SadTalker, and SyncTalk. In \emph{Overall Naturalness} it likewise achieves the highest mean score of $3.73\pm1.07$, significantly outperforming DIM, SyncTalk, and SadTalker. Every comparison is significant on both dimensions, with effect sizes $d=0.45$ to $1.49$, so even the weakest comparison exceeds the conventional small-effect threshold. This is consistent with our advantage on Interaction DTW and shows that our method balances individual animation realism with dyadic coordination rather than trading one for the other.

The results of the forced-choice preference test further confirm the perceptual superiority of our method. Among all 112 valid choices, DiaHeads is selected as the most natural animation in 51.79\% of cases, while SadTalker, DIM, and SyncTalk account for 29.46\%, 15.18\%, and 3.57\% respectively. All DiaHeads pairwise comparisons are significant under two-proportion $Z$-tests ($Z=3.40$ to $8.06$, $p\le0.004$). The only non-significant pair is SadTalker versus DIM ($p=0.061$), which is consistent with both being substantially behind our method. Participants thus consistently rate DiaHeads as having better overall realism and significantly superior interactional coordination compared to state-of-the-art baselines, which addresses the core requirement of generating natural, socially coherent dialogue animations for offline content creation.

\section{Conclusion}
\label{sec:conclusion}

We presented DiaHeads, a framework that jointly generates 3D head animations for both participants from their audio streams, treating the two speakers on equal terms without role conditioning. This approach directly models the reciprocal and coordinated nature of non-verbal interaction, including gaze, head motion, and expressions, and it carries an explicit, per-frame, four-cue discrete representation for both participants. We also introduced the CANDOR-FG dataset, extending CANDOR with synchronized FLAME parameters and gaze directions for dyadic dialogues, together with an analysis of the reliability of those automatically estimated targets. Experiments covering coordination diagnostics, two ablation studies, and a statistically analyzed human subject evaluation show that DiaHeads outperforms existing baselines in producing more coordinated and natural interactions. The generated animations target offline content creation, such as animating non-player-character conversations and pre-visualizing film sequences.

CANDOR-FG also supports work beyond dyadic dialogue animation. Its synchronized FLAME and gaze data can be used for real-time interactive avatars and for data-driven models of co-speech gaze and head motion in social robotics.

\balance

\bibliographystyle{IEEEtran}
{\footnotesize
\bibliography{refs}
}

\end{document}